\documentclass[aps, twocolumn, showpacs]{revtex4}

\begin{document}
\title{Lorentz covariance and gauge invariance in the proton spin problem}
\author{S. C. Tiwari \\
Department of Physics, Banaras Hindu University, Varanasi 221005, \\ Institute of Natural Philosophy, 
Varanasi India\\}
\begin{abstract}
In this brief note insightful remarks are made on the controversy on the decomposition of the proton spin into the spin and orbital angular momenta of quarks and gluons. It is argued that the difference in the perception on the nature of the problem is the main reason for the persistent disputes. There is no decomposition that simultaneously satisfies the twin principles of manifest Lorentz covariance and gauge invariance, and partial considerations hide likely inconsistencies. It is suggested that field equations and matter (i. e. electron in QED and quarks in QCD) equations must be analyzed afresh rather than beginning with the expressions of total angular momentum; canonical or otherwise.
\end{abstract}
\maketitle

\section{\bf Introduction}

Deep inelastic scattering (DIS) electron-proton experiments established point-like internal structure of proton in which the constituents (partons) seemed to be noninteracting free particles. Infinite momentum frame (IMF) is widely used to understand DIS experiments. Quark hypothesis leads to quantum chromodyamics (QCD) where quark-quark interaction is mediated by strong gauge fields - gluons. Physical interpretation of scattering experiments has two approaches: phenomenological parton model and relativistic kinematics, and perturbative QCD. It has to be emphasized that gauge and relativistic invariances have to be maintained even in the perturbative approximation. Polarized DIS experiments challenged the theoretical understanding on the internal spin structure of the proton that has come to be known as proton spin puzzle. Experimental ingenuity to extract information on the distribution of proton spin amongst its constituents is being tested world-over \cite{1}. At a fundamental level the question is that of separating proton spin into spin angular momentum (SAM) and orbital angular momentum (OAM) of quarks and gluons. Recent interest and intense debate have the origin in the exaggerated claim \cite{2} that 'the long standing gauge invariance problem of the nucleon spin structure' has been solved. Lorce made an attempt \cite{3} to reconcile the disputes arising out of the various spin decompositions existing in the literature by distinguishing active and passive gauge transformations.

It seems there is a basic problem at the level of the understanding as to what the nature of the problem is. My understanding \cite{4} is that simultaneous manifest Lorentz covariance and gauge invariance have not been achieved either in \cite{2} or in any other formulation. This, of course, does not disqualify a particular decomposition from its practical utility to correlate experimental data with physics: additional assumptions consistent with the experimental situation are natural. Recent reviews \cite{5, 6}, though informative and educative, do not resolve the nucleon spin controversy satisfactorily. In this note I propose to offer new insights on the symmetry and invariance relevant to the issue. Electrodynamics serves important role in exploring QCD \cite{7, 8, 9}. In the next section salient features of classical electrodynamics (CED), quantum electrodynamics (QED) and quantum optics are revisited. Proton spin puzzle is discussed in Section III, and the last section ends with speculative remarks.

\section{\bf Fields, potentials and sources}

We focus on observables and measurement, Lorentz covariance and relativistic invariance, and gauge invariance. Electromagnetic (EM) fields are observable but not the EM potentials. What does this statement mean? Do we really measure EM fields? In the mechanical world-view, force being something that is directly related with the sense experience, we tend to believe in its physical reality. However, primarily the measurement is that of spatial correlation and counting on a calibrated meter scale. Theoretical construct is essential to relate the meter reading with a physical quantity, e. g. force. An abstraction has to be made for electric field: limit of the Coulomb force divided by test charge tending to zero. One is free to seek description of CED dispensing with the fields altogether, for example, in action-at-a-distance theories. Even in field theory independent reality to EM fields crucially depends on the energy and momentum carried by the so called radiation fields that are assumed to have separated from the sources; here boundary conditions become very important. It is reasonable to state that CED in vacuum comprises of Maxwell field equations and Newton-Lorentz equation of motion for the charges with appropriate boundary conditions.

One may introduce EM potentials as mathematical tools for convenience; gauge transformation is just a mathematical curiosity in that case. Relativistic invariance also has contextual merit: static fields and slowly moving charged particles in external EM fields are nice approximations to many physically realizable situations. Relativistic electrodynamics could be studied adopting Einstein's original algebraic approach for the Lorentz transformations. Thus manifest Lorentz covariance and gauge invariance are not fundamental issues in this formalism of CED; Section 15-6 in Feynman lectures \cite{10} is a typical misleading commentary where time-independent electromagnetism is termed as false rather than an extremely  useful approximation to CED. In fact, we know today that electron not only has electric charge it has spin, and it has self-field, therefore from rigorous point of view CED cannot be accepted as a complete description of EM phenomena even at a classical level.

Standard relativistic field theories begin with postulating an action functional and variational principle; symmetries of the dynamical system correspond to that of the action. A thought provoking presentation can be found in Corson's book \cite{11} and Eddington's monograph \cite{12}; see also \cite{13}. Consider relativistic invariance. The Lagrangian density $L$ is Lorentz invariant and its spacetime or 4-volume integral is obviously invariant. Assuming a constant time surface amounts to the specification of a Lorentz frame, however the Euler-Lagrange equations derived in this way are Lorentz covariant. It is possible to replace constant time surfaces by space-like surfaces maintaining manifest relativistic invariance in the variational procedure; see Section 19.a in \cite{11}.

Recall that relativistic invariance can be considered algebraically applying Lorentz transformations on physical quantities in different inertial frames. Manifest Lorentz covariance makes sense in geometrical approach postulating 4-dimensional (4D) spacetime continuum following Minkowski. Maxwell equations written in terms of the pair of vector fields $({\bf E},{\bf B})$ are relativistically invariant, but these fields do not satisfy vector or tensor transformation laws in 4D spacetime geometry. In the well known construction these fields are combined to obtain antisymmetric EM field 4-tensor $F_{\mu\nu}$ and the Maxwell equations get transformed to the covariant form. Most often, covariance of equations is of great utility, however strict adherence to this principle by itself may lead to unphysical situations since arbitrary new auxiliary fields could render any noncovariant equation into a covariant form. Trautman gives an example \cite{14} of a general covariant form
\begin{equation}
 u^\mu A_\mu =0
\end{equation}
of the equation
\begin{equation}
 A_1 =0
\end{equation}
introducing the coordinate basis vector field $u^\mu$.

Usually symmetry of the action $S$ is also a symmetry of the equations of motion, but there do exist exceptions, e. g. duality rotation in EM theory. In addition to this the formalism based on the action principle may admit new characteristics. We have pointed out that EM potentials $A_\mu$ are superfluous in CED, but in the variational procedure they acquire fundamental role as independent field variables. The most remarkable consequence is that the invariance of the action $S_M$ for free EM fields under the gauge transformation
\begin{equation}
 A_\mu \rightarrow A_\mu + \partial_\mu \chi
\end{equation}
introduces a new symmetry. Note that in the Maxwell field equations there is no need to introduce EM potentials, and hence to discuss gauge invariance. Not only this, even the standard arguments overlook a significant point that seems to have not been discussed in the literature. The set of homogeneous Maxwell equations
\begin{equation}
 {\bf \nabla}.{\bf B} =0
\end{equation}
\begin{equation}
 {\bf \nabla} \times {\bf E} = - \frac{\partial {\bf B}}{\partial t}
\end{equation}
are shown to be equivalent to the following in the textbook approach, Section 6.4 in \cite{15}
\begin{equation}
 {\bf B} ={\bf \nabla} \times {\bf A}
\end{equation}
\begin{equation}
 {\bf E} =- {\bf \nabla}\Phi - \frac{\partial {\bf A}}{\partial t}
\end{equation}
Unfortunately expression (6) as a consequence of the vanishing of the divergence of ${\bf B}$ holds only in 3D space; for time-varying fields (6) has to be generalized to 
\begin{equation}
 {\bf B} ={\bf \nabla} \times {\bf A} + {\bf V}(t)
\end{equation}
Obviously using (8) in Eq.(5) is inconsistent with the expression (7). There is an important physical implication on the relativistic invariance: time variable in one inertial frame becomes space coordinate dependent in another inertial frame and the Lorentz transformation of (8) is complicated due to time-dependent vector ${\bf V}(t)$. It will have bearing on the question of frame-dependence discussed in \cite{5, 6}. The covariant form of (6) and (7) given by the following expression defines EM field tensor
\begin{equation}
 F_{\mu\nu} = \partial_\mu A_\nu - \partial_ \nu A_\mu
\end{equation}
However in the light of (8) this elegant form can no longer be maintained. Citing the earlier literature Leader \cite{8} and Leader and Lorce \cite{6} argue that $A_\mu$ is not a Lorentz 4-vector but has the transformation law (Eq. 202 in \cite{6})
\begin{equation}
 A^\mu (x) \rightarrow  \Lambda^\mu_\nu (A^\nu(x) + \partial^\nu \omega_\Lambda(x))
\end{equation}
Now even if we discard ${\bf V}(t)$ and assume (9) the relativistic transformation law (10) inclusive of gauge transformation needs careful consideration. In principle, there is no objection to postulate wider group of transformations. One of the earliest generalizations to the general coordinate transformations of general relativity is that of Weyl's unified theory in which the gauge transformations were introduced for the first time which later changed to phase transformations for compatibility with quantum theory. In the Weyl geometry EM potentials and fields acquire geometrical significance \cite{16}. We refer to Dirac's concise presentation \cite{17} on the rudiments of Weyl geometry: one defines in-invariants, in-tensors and co-tensors depending on the gauge transformations (or Weyl powers) and coordinate transformations in a consistent manner. Lorce's generalized Lorentz transformations (GLT) (10) mix gauge and Lorentz transformations. Do they have universal applicability? Are they applicable only to EM potantials? Note that the Lorentz condition
\begin{equation}
 \partial_\mu A^\mu =0
\end{equation}
is similar to the current continuity equation
\begin{equation}
\partial_\mu J^\mu =0 
\end{equation}
Does $J^\mu$ follow the transformation rule (10)? In the absence of a consistent set of rules for GLT to classify various physical quantities the transformation (10) is nothing more than a mathematical artefact. Physically, interwining of gauge and Lorentz transformation would imply nonequivalence of inertial frames contrary to the spirit of special relativity.

Similarity of the mathematical structure of Eqs. (11) and (12) is significant but physical interpretation differs. Consider (12) for which the solution has the form
\begin{equation}
 J_\mu = \partial^\nu F_{\mu\nu}
\end{equation}
where $F_{\mu\nu}$ is antisymmetric tensor. Since the solution (13) is not unique there is a freedom of gauge transformation
\begin{equation}
 F_{\mu\nu} \rightarrow F_{\mu\nu} + M_{\mu\nu}
\end{equation}
with $M_{\mu\nu}$ being another antisymmetric tensor. If $F_{\mu\nu}$ is identified with EM field tensor then Eq.(13) is just the Maxwell field equation. However in view of (14) the Maxwell field has arbitrariness. A detailed analysis on Maxwell-Lorentz theory and Maxwell-Dirac theory \cite{18} shows that classical EM field tensor hides spin tensor $M_{\mu\nu}$, i. e. even in the usual source-free case the EM fields are not decoupled from the sources. It has been recognized that the question of AM of the electron and photon is intricate in interacting QED. Leader in Section VI F of \cite{8} compares Chen et al decomposition with that of Wakamatsu and argues that the later is merely a rearrangement of the former and could be done in infinite number of ways. Note that following the work of Chen et al it has become customary to begin the discussion with the chosen expression of the total AM integral. I think the deeper issue of the source-field duality and the equations of motion is crucial to have a complete picture. Leader's arguments are unsatisfactory considering the gauge transformation (14) and the likely new approach to interacting QED.

In fact above line of thought could be continued adopting Weyl's original theory. In the generalized Weyl-Dirac theory \cite{19} the current density is found to be
\begin{equation}
 J^\mu = \partial^\mu \xi + 2 \xi A^\mu
\end{equation}
Here $\xi$ is a co-scalar field in Weyl geometry. A beautiful aspect of Eq.(15) is that three distinct cases can be consistently envisaged with interesting physics. (I) $J^\mu =0$ implies $A^\mu$ is a pure gauge field and EM field tensor $F^{\mu\nu}$ vanishes. (II) $\xi =0$ implies $J^\mu =0$, however $A^\mu$ need not be zero. Thus the field is decoupled from the source. (III) $\xi = constant$, then (15) resembles relativistic London equation in superconductivity.

We have seen that already at the classical level the field-matter interaction is not simple, and there are unresolved fundamental problems. In quantum theory additional complications arise due to the combination of relativity, quantum rules and gauge invariance. Leader in the expository article \cite{8} puts the spin decomposition controversy in perspective revisiting QED; I mention two of the several interesting points made by him. First he notes that in the papers on the controversy 'the treatment is essentially classical'. It is true, and in my view its implication should be that the understanding at the classical level itself has confusions. In the short comment \cite{4} this point was emphasized both for CED and optics experiments; I refer to third paragraph on SAM and OAM of light beams. It may be further added that Lax et al \cite{20} brought out an apparent paradox, 'Experimentally, the laser-oscillator modes found in the apparently inconsistent way agree extremely well with those predicted by this theory'. The theory refers to paraxial wave optics that is at the basis of the seminal work on OAM by Allen et al \cite{21}. A curious gauge theoretic formulation based on complex scalar field could be found in \cite{22}. 

Could spin of photon be converted to photon OAM? If spin is an intrinsic property of photon then the claim of SAM to OAM conversion defies satisfactory explanation. What is the relationship of polarization with spin? Does helicity description resolve the problems? There are a large number of papers on these questions differing on physics of the observed phenomena. If, following Poynting we assume AM as the integral of ${\bf r} \times({\bf E} \times{\bf B})$ and calculate it as $\frac{\lambda}{4\pi}$ times the linear momentum of the polarized wave then Beth experiment \cite{23} can be easily explained in which polarized light beam exerts a torque on a doubly refracting plate. Photon concept, gauge invariance and separation of AM into SAM and OAM are nonissues. It is for these reasons that many statements in \cite{2} are misleading, e. g. ``Beth made a direct measurement of the photon spin', and also remarks referring to OAM citing \cite{21} and \cite{24}. 

Leader's second point is that in QED one measures matrix elements of the operators and the role of physical states in the Hilbert space becomes important, therefore gauge invariance of the operators has to be seen with reservations. Based on the Coulomb gauge QED he argues that $A_\mu$ is not a true Lorentz 4-vector; Lorce takes this view further to suggest GLT \cite{6}. In the classical context we have pointed out above that GLT lacks a clear foundation. In QED we draw attention to the literature having bearing on this issue. Schwinger in his Nobel lecture \cite{25} makes important points: 'microscopic measurement has no meaning apart from a theory', 'measurements individually associated with different regions in space-like relation are causally independent, or compatible', and explains relativistic invariance of the action principle using the invariant 3D surface in the space-like relation. Note Corson's remarks in this connection mentioned earlier \cite{11}. The meaning of time variable in quantized theory thus has debatable ramification on covariance in QED. Haller also made important contributions on covariant gauge in QED \cite{26}.

In the context of observables and Coulomb interaction in QED I invite attention to a transparent and thoughtful article by Schweber in \cite{27}. It is worth quoting from this article: ''The Coulomb gauge theory and the Lorentz gauge theory thus both describe the same physical phenomena, but they handle one aspect of the physical situation, namely the Coulomb interaction, in fundamentally different ways. In the Coulomb gauge the interaction is incorporated into the electron field, while in the Lorentz gauge it appears as being caused by the emission and absorption of longitudinal quanta.`` It seems the separation of momentum in QED in \cite{7, 8} could be viewed afresh in the light of the above quotation. Section 11.2 on Invariance Properties in \cite{27} and the remarks at the end of this section on the Lorentz transformation of $A_\mu$ are quite illuminating; together with the paper by Strocchi and Wightman \cite{28}  also cited by Leader \cite{8} would be useful to understand QED better. I think ESAB effect, now termed as AB effect, is not discussed convincingly in \cite{28}. Note that ESAB effect could be nicely understood as topological effect without invoking quantum theory; Wilson lines and loops essentially generalize this concept. Therefore, for nontrivial, e. g. multivalued pure gauge potentials there could be observable topological effects. In the next section we return to this point.

\section{\bf Remarks on the proton spin controversy}

The proposed decomposition of total AM of nucleons into spin and orbital parts of quarks and gluons in \cite{2} has been elucidated in a series of papers, amogst others by Wakamatsu, Leader and Lorce; however it seems puzzling to find disagreements between them on the nature of the basic issues. Preceding discussion focused on electromagnetism helps us delineate some of the questions. I also invite attention of the QCD community to the insightful discussions given in Sections 19 and 22 of \cite{11} referred to in \cite{4}. Note that the starting point of \cite{2} is the conserved Noether canonical QCD AM tensor; though Corson does not use the term Noether's theorem the treatment in \cite{11} essentially corresponds to this.

After devoting considerable time to understand the viewpoints expressed in \cite{2, 3, 5, 6, 7, 8, 29, 30, 31} I realize that the main reason for disagreements is that the twin principles of Lorentz covariance (LC) and gauge invariance (GI) have markedly different priorities and perceptions, as well as ambiguities in these works. To begin with, a careful reading of \cite{2} shows that the principal idea of the authors is to split the gauge potential into a pure gauge part and a so called physical part
\begin{equation}
 A_\mu^a = A_\mu ^{a, pure} + A_\mu ^{a. phys}
\end{equation}
Gauge potential, $A_\mu^a$, by definition, is a Lorentz 4-vector with $\mu$ Lorentz group index, and SU(3) color index, a. Under the gauge transformation it transforms as
\begin{equation}
 A_\mu^a  \rightarrow U[A_\mu^a + \frac{i}{g} \partial_\mu]U^\dagger
\end{equation}
Decomposition (16) is motivated by the following gauge symmetry prescription
\begin{equation}
 A_\mu^{a, pure}  \rightarrow U[A_\mu^{a, pure} + \frac{i}{g} \partial_\mu]U^\dagger
\end{equation}
\begin{equation}
 A_\mu^{a, phys}  \rightarrow U A_\mu^{a, phys} U^\dagger
\end{equation}
Chen et al rightly emphasize that how to determine these gauge potentials is a nontrivial task. In the simpler case of QED with U(1) gauge group the transformations (18) and (19) reduce to
\begin{equation}
 A_\mu^{pure} \rightarrow A_\mu^{pure} + \frac{1}{g} \partial_\mu \alpha
\end{equation}
\begin{equation}
 A_\mu^{phys} \rightarrow A_\mu^{phys}
\end{equation}
The space components of the potentials are determined by
\begin{equation}
 {\bf \nabla}.{\bf A}^{phys} = 0
\end{equation}
\begin{equation}
 {\bf \nabla} \times {\bf A}^{pure}=0
\end{equation}
Incidentally the defining equations (22) and (23) are same as those satisfied by the transverse and the longitudinal components of the vector potential
\begin{equation}
 {\bf \nabla}.{\bf A}^{tr} = 0
\end{equation}
\begin{equation}
 {\bf \nabla} \times {\bf A}^{long}=0
\end{equation}
Let it be emphasized that the decomposition of vector potential into pure and physical parts originates in GI, and into transverse and longitudinal parts has meaning in the context of inertial frames. The implication of the statement in \cite{2} that,'These are nothing but the transverse and longitudinal components of the vector potential ${\bf A}$' is misleading since indistinctness between defining expressions (22)-(23) and (24)-(25) introduces ambiguity in physical interpretation. 

What are the corresponding relations in QCD? Eqs. (14) and (18) in their paper \cite{2} are the proposed relations; the first one is obtained as a generalization to the curl-free condition replacing ${\bf \nabla}$ by pure gauge derivative ${\bf D}^{pure}$ while the other relation is based on the imposition of gauge invariance on the gluon OAM in their proton spin decomposition. Recall that in the Lagrangian field theory $A^a_\mu$ are fundamental independent field variables and conserved AM is a consequence of the rotational symmetry of the Lagrangian or action. Reversing the logic to constrain $A^a_\mu$ from OAM seems a flawed approach.

To summarize the inherent weaknesses in Chen et al approach: W1) GI dominates the basic new idea in the noncovariant form of the splitting (16), however the transverse-longitudinal separation is interchangeably invoked confusing the roles of GI and LC, W2) decomposition of canonical AM has inbuilt arbitrariness, and W3) the constraint on physical component of gluon vector potential is ad hoc demanding gauge invariance of gluon OAM.

In the light of above commentary we offer following remarks.

${\bf R1}$: Much of the proton spin controversy is germinated in W1. In the perception of Wakamatsu and Ji  transverse-longitudinal paradigm dominates the splitting of $A^a_\mu$, while Leader and Lorce are willing to mix LC and GI. Arguments built on these different perceptions lead to differing physical interpretations. The notion of observables and uniqueness of the decomposition have been given different meanings. Wakamatsu argues \cite{29} that the question of uniqueness considered by Lorce \cite{30} invoking so called Stueckelberg symmetry ignores the important role of transversality; he is also critical of the idea of gauge-invariant extension (GIE). In this connection Wakamatsu's statements that, 'Note that we are using the word ''observables`` in a strict sense. That is, they must be quantities, which can be extracted purely experimentlly , i. e. without recourse to particular theoretical schemes or models' amount to just the collection of empirical data not physical quantities. We refer to the first paragraph in the preceding section and also recall the quoted Schwinger's remarks. Wakamatsu asserts that 'the choice of gauge, the choice of Lorentz frame, and the transverse-longitudinal decomposition are all intrinsically interwined'; I am unable to agree with this because such a viewpoint is also responsible for the confusion between LC and GI.

Lorce proposes two notions of gauge invariance termed as passive and active gauge transformations inspired by general relativity \cite{3}. Stueckelberg transformation is suggested to be a combination of the two. He seeks to 'avoid confusion' using the terminology of observables and quasi-observables, however both are measurable in the sense that 'they can in principle be extracted from experimental data'. If both are measurable, physically their difference is trivial and inconsequential. Regarding the Stueckelberg symmetry, Wakamatsu's criticism \cite{29, 32}, at least, in electromagnetism seems valid.

The paper by Ji et al \cite{31} is interesting in the sense that LC and GI are not only interwined the meaning of Chen et al decomposition and Lorentz transformations are altered: the gluon spin operator ${\bf E}^a \times {\bf A}^a_{phys}$ in the expresssion (12) of \cite{2} is replaced with ${\bf E}^a \times {\bf A}^a_{tr}$ (see second paragraph in \cite{31}), and it is stated that 'the frame dependence cannot be obtained from Lorentz transformation and is a function of dynamic details'. Though IMF limit plays important role in Ji et al paper, it is not obvious to accept their assertion that Wakamatsu's statement about the frame independence is incorrect. Authors do not explain how dynamical frame dependence differs from GLT of Lorce and Leader.

${\bf R2}$: Lorce has pointed out primarily four types of proton spin decomposition in the literature, namely Jaffe-Manohar \cite{33}, Ji \cite{34}, Chen et al \cite{2} and Wakamatsu \cite{7}. I think prior to the recent controversy that began in 2008 the construction of total AM and its separation into SAM and OAM of quarks and gluons had reasonable theoretical and experimental foundations. For example, in the interacting gauge theory the improved form of the Belinfante symmetric and gauge invariant energy-momentum tensor is used to construct the AM tensor by Ji \cite{34}. Gauge invariance guides him to restrict the decomposition to  quark SAM, quark OAM and gluon AM since gluon AM cannot be separated into spin and orbital parts gauge invariantly.

Leader highlighted \cite{8} the arbitrariness in the proton spin decomposition as having 'infinite number of possibilities' in \cite{7} that was based on the proposition of Chen et al \cite{2}. The review \cite{6} also concludes that there are infinity of the decompositions contrary to the belief of Wakamatsu \cite{29, 32} that there exist only two physically in-equivalent gauge invariant decompositions. We have argued \cite{35} that topological ideas could throw light on this issue. The contentious point is that of GIE. We reproduce few sentences from the review \cite{6}.

${\bf GIE-LL}$: Consider any gauge noninvariant object in a specific gauge, one can easily construct a gauge-invariant object leading to the same physical result in any gauge. Formally, it is sufficient to replace the full gauge potential $A^\mu$ by the physical field $A_\mu^{phys}$ and the ordinary derivative $\partial_\mu$ by the appropriate pure-gauge covariant derivative. Starting from a single gauge non-invariant local quantity, one can construct infinitely many GIEs.

According to \cite{6} Wakamatsu seems to have overlooked the difference between the gauge-fixing procedure and the Stueckelberg-fixing procedure, and 'wrongly concluded that all the GIEs arte physically equivalent'. Leader and Lorce introduce a new term which they call as Lorentz-invariant extension (LIE): the Lorentz invariant quantity $p^\mu p_\mu$ is a LIE of the quantity $m^2$, which agrees with the value of $m^2$ in the rest frame. The measurement is explained invoking LIE since the rest mass is never measured in the high energy scattering experiments. It is surprising that the authors uncritically assume $m=\gamma m_0$ since in relativity it is not the rest mass that is boosted in an inertial frame with relative velocity $v$ but it is the energy-momentum 4-vector that is Lorentz transformed from one frame to the other. Note that at a fundamental level, the notion of rest mass has inherent circularity (e. g. Mach's criticism), and in principle cannot be measured even in nonrelativistic case \cite{36}. The imprint of the weakness W1 is quite evident in the proposed LIE-GIE similarity; a critical evaluation of LIE and GIE is necessary.

${\bf R3}$: Initially the authors \cite{2} assumed the constraint on physical vector potential to be that its commutator with ${\bf E}^a$ vanishes.
Subsequently it was changed \cite{37} to ${\bf D}^{a, pure}. {\bf A}^{a, phys}=0$. Lorce \cite{38} interprets this generalized Coulomb constraint in terms of Stueckelberg symmetry breaking. He says that one is free to use any other constraint, for example, the light-front constraint. I think in the absence of a sound mathematical or physical reasoning the ad hoc character of the defining relation on the physical part of gluon vector potential remains one of the most unsatisfactory aspects of Chen et al approach.

${\bf R4}$: Personal views and subjective interpretations have dominated the literature on proton spin controversy since 2008. Twisting or partial modifications in the standard physics on LC and GI cannot be accepted. Either we strictly follow the standard physics or explore radically new ideas. In the next section I share some of my unorthodox ideas, however let us first follow the established understanding to address some of the questions.

(i) General relativity as a gauge theory inspired by nonabelian Yang-Mills theory has undoubtedly been a fascinating subject. It is also known that the solution of Einstein equation is unique only up to a diffeomorphism; 4-gauge conditions on the covariant derivative of the metric tensor with respect to the background metric are similar to the Lorentz gauge condition in electrodynamics \cite{39}. In spite of this, general relativity is fundamentally different since the spacetime is dynamical, and the notion of local frame and the coordinate basic vector fields help make only superficial similarity with the local gauge transformations. Note that Weyl's gauge group is a noncompact group of homothetic transformations, and modern gauge theories, in contrast, are based on compact U(1)-bundle or SU(n)-bundles. Let it be re-emphasized that the spacetime in QED and QCD is nondynamical flat Minkowskian.

(ii) In the light of Eq.(8) the EM field tensor (9) does not follow from the Maxwell equations unless we set ${\bf V}(t)=0$ as initial condition that holds for all time. On the other hand we begin with the postulate that $A^\mu$ is a 4-vector in the variational formulation, construct $F^{\mu\nu}$ to define the Lagrangian density for the Maxwell field to be
\begin{equation}
 L_M = -\frac{1}{4\pi} F^{\mu\nu} F_{\mu\nu}
\end{equation}
Thus demanding proof that $A^\mu$ is a 4-vector from Maxwell equations is not a right question. Moreover, GLT (10) has a conventional interpretation: $\omega_\Lambda$ is a Lorentz scalar $\rightarrow$ $\partial ^\mu \omega_\Lambda$ transforms as a Lorentz 4-vector $\rightarrow$ $A^\mu$ also transforms as a 4-vector.

(iii) Covariance in the geometric approach to special relativity strictly implies relativistic invariance. Thus the Lorentz gauge (11) is manifestly Lorentz covariant. Is it also gauge invariant? No. Under the gauge transformation (3) Eq.(11) becomes
\begin{equation}
 \partial_\mu A^\mu + \partial_\mu \partial^\mu \chi=0
\end{equation}
Gauge-fixing requires addition of $B\partial_\mu A^\mu$ term in $L_M$ where $B(x)$ is a Lagrange multiplier. Thus the action with gauge-fixing term is Lorentz invariant but gauge noninvariant unless we impose the condition
\begin{equation}
 \partial_\mu \partial^\mu \chi =0
\end{equation}
Coulomb gauge is neither Lorentz covariant nor gauge invariant.

(iv) Trautman's example, Eqs.(1) and (2), shows that any noncovariant form could be made covariant using appropriate auxiliary fields in general relativity; note the role of $u^\mu$ in Eq.(1). In flat spacetime there will be a kind of artificiality, however Wakamatsu \cite{29, 32} constructs covariant gauge-fixing condition
\begin{equation}
 n^\mu A_\mu =0
\end{equation}
Eq.(29) represents a class of gauges: the temporal, the light-cone and the spatial-axial gauges defined respectively by $n^\mu = (1, 0, 0, 0), \frac{1}{\sqrt 2} (1, 0, 0, 1), and ~ (0, 0, 0, 1)$. Do covariant gauges (29) interwine LC and GI? The choice of $n^\mu$  characterizes the Lorentz frame of the gauge-fixing. In special relativity all inertial frames are equivalent. Clearly gauge-fixing should not be allowed to interfere with this.

(v) Trautman's argument shows that auxiliary fields may turn out to be unphysical in some situations \cite{14} and this defect seems to be echoed in GIE-LL. Such an idea as that of GIE could, at best have a limited value in specific instances, but it is not acceptable as a sort of a principle.

(vi) Nonlocal effect of gauge potential in ESAB effect is well known; Wilson line is a generalization of the AB phase integral. One usually specifies a path along which the gauge potential is evaluated, thus nonlocal Wilson lines are path-dependent. Experiments demonstrate AB effect in closed loops: shift in the interference pattern of two electron beams encircling a solenoid. Path-dependence of the nonlocal effect has not been proved experimentally for open paths. In QCD the choice of a gauge, for example, light-cone gauge makes it difficult to visualize Wilson loops. If nonlocal path-dependent gauge potential effect gets compensated by a choice of the gauge the uniqueness claim of Wakamatsu seems justified. In that case only for multiply connected spacetime, a nontrivial topological effect could exist and infinity of decompositions could be envisaged \cite{35}.

(vii) A consistent and complete theory starting from the first principles incorporating (16) does not exist; the only welcome exception is an attempt by Lorce \cite{38} in this direction. Unfortunately the claim that Chen et al decomposition is derived from 'the standard procedure based on the Noether's theorem' is not correct. The reason is that there are two ways to incorporate constraints in the variational approach that in general give different results. Rigorous procedure is to incorporate constraints through Lagrange multipliers. Lorce makes use of the constraint on pure gauge field (his equation (20)) in the construction of the Lagrangian density itself and hence obtains the results in conformity with Chen et al. Let me explain the rigorous procedure in QED example. Decomposition (16) leads to two field tensors $F^{\mu\nu , pure}$ and $F^{\mu\nu, phys}$ and $L_M$ is proportional to
\begin{equation}
 F^{\mu\nu}F_{\mu\nu} = F_{\mu\nu}^{pure}F^{\mu\nu, pure} + F_{\mu\nu}^{phys}F^{\mu\nu, phys} + 2 F_{\mu\nu}^{pure}F^{\mu\nu, phys}
\end{equation}
Imposing the constraint
\begin{equation}
 F^{\mu\nu, pure} =0
\end{equation}
an additional term arises in the Lagrangian density proportional to
\begin{equation}
 L_1 = C_{\mu\nu} F^{\mu\nu, pure}
\end{equation}
In the variational procedure both $A^{pure}_\mu$ and $A^{phys}_\mu$ have to be treated as independent field variables. Obviously the Euler-Lagrange equations and the conserved quantities obtained in this way may differ from those obtained by Lorce in QCD. One well known fact has also to be reiterated regarding quark field equations: Dirac current is independent of the gauge field even under local gauge transformations. It is suggested that a model in which quark field in QCD is replaced by a Lorentz scalar field could throw light on some of the issues; the work on these problems is under progress, here only intuitive expectations have been pointed out.

\section{\bf Conclusion}
In this contribution a critique on the conceptual issues in the proton spin controversy has been presented keeping in mind the spirit of the Workshop. Apart from the prejudices the lack of decisive direct experiments on the measurement of SAM and OAM of quarks and gluons is mainly responsible for the unending disputes. The fall out of Chen et al proposal could be seen in two different ways. Firstly it has inspired re-examination of subtle aspects on gauge symmetry and Lorentz covariance in QED and QCD. Second one is a negative impact on the efforts to resolve proton spin puzzle. The introductory paragraph in \cite{2} notes 'disturbingly and surprisingly' that even after 20 years of the puzzle the first task of the proton spin decomposition into SAM and OAM of quarks and gluons has not been done. Why? One of the reasons could be that this task seems almost irrelevant from practical point of view; the recent comprehensive review \cite{1} only passingly mentions it. Experimentally the theory that directly relates with various kinds of parton distributions and their moments, and has utility in the factorization theorems is important \cite{1, 40, 41, 42}. For example, gluon helicity distribution has immense interest \cite{31, 43}. I think the insights developed in the present work could prove useful in addressing some of the controversial issues. Topological ideas and point (vii) in the remark R4 have the potential to throw light on the proton spin puzzle.

I conclude the article with my speculations having bearing on this problem. In a radically new interpretation \cite{44} photon is proposed to be a topological defect such that ${\bf E} = {\bf B} =0$, and rotating photon fluid is described by EM field tensor as its AM tensor. Further the new hypothesis explored over past several years is that rotating spacetime is fundamental and the origin of charge is attributed to (fractional) spin $\frac{e^2}{c}$ \cite{45}. QCD hypothesis that free quarks do not exist is replaced by the idea that neutrinos confined inside the hadrons behave as quarks. I believe the meaning and measurement of spin for nucleons will play a pivotal role in assessing such speculations.

${\bf Acknowledgments}$  I am grateful to Mauro Anselmino for invitation to the ECT Workshop, and agreeing to circulate this article to the participants. I am grateful to M. Wakamatsu for his continued interest in my work, and thank E. Leader and C. Lorce for useful correspondence on this controversy.

\end{document}